\documentclass[aps, superscriptaddress, twocolumn,prl,floatfix]{revtex4}
\usepackage{amssymb}
\usepackage{amsmath}
\usepackage{graphicx}
\usepackage{color}
\usepackage{natbib}

\def\Gq{e^2/h}

\def\Vtg{V_{TG}}
\def\Vbg{V_{BG}}
\def\Ctg{C_{TG}}
\def\Cbg{C_{BG}}
\def\sigmanp{\mathrm{\sigma}_{CNP}}
\def\gnp{g_{CNP}}
\def\rnp{r_{CNP}}
\def\dEf{\delta E_{F}}

\def\mueV{\mu eV}

\begin{document}
\title{Insulating behavior at the neutrality point in single-layer graphene}
\author{F. Amet}
\affiliation{Department of Applied Physics, Stanford University, Stanford, CA 94305, USA}
\author{J. R. Williams}
\affiliation{Department of Physics, Stanford University, Stanford, CA 94305, USA}
\author{K.Watanabe}
\affiliation{Advanced Materials Laboratory, National Institute for Materials Science, 1-1 Namiki, Tsukuba, 305-0044, Japan}
\author{T.Taniguchi}
\affiliation{Advanced Materials Laboratory, National Institute for Materials Science, 1-1 Namiki, Tsukuba, 305-0044, Japan}
\author{D. Goldhaber-Gordon}
\affiliation{Department of Physics, Stanford University, Stanford, CA 94305, USA}

\begin{abstract}
The fate of the low-temperature conductance at the charge-neutrality (Dirac) point in a single sheet of graphene is investigated down to 20\,mK.  As the temperature is lowered, the peak resistivity diverges with a power-law behavior and becomes as high as several Megohms per square at the lowest temperature, in contrast with the commonly observed saturation of the conductivity. As a perpendicular magnetic field is applied, our device remains insulating and directly transitions to the broken-valley-symmetry, $\nu$=0 quantum Hall state, indicating that the insulating behavior we observe at zero magnetic field is a result of broken valley symmetry. Finally we discuss the possible origins of this effect.
\end{abstract}
\maketitle

The ability to create electronic devices in graphene has made it possible to study 2D Dirac fermions in the solid state~\cite{Geim2007}.  Transport measurements in a large magnetic field display quantum Hall plateaus with unconventional values of conductance, a signature of the Dirac equation describing electrons in graphene~\cite{Geim2007, Zhang2005}. When the cyclotron gap becomes larger than disorder-induced fluctuations in the surrounding potential, the effect of the linear Dirac band structure becomes evident. At zero magnetic field, the disorder landscape dominates~\cite{Martin2008}, blurring the interesting phenomena that might occur at the Dirac point. Recently, the influence of disorder has been reduced by either suspending a sheet~\cite{Bolotin2008} or placing it on atomically-flat boron nitride (BN)~\cite{Dean2009}, and many discoveries in transport have been made due to the more readily-accessible Dirac point in these cleaner system~\cite{Elias2011,Bolotin2009,Dean2011}. Anomalous patterns in the magneto-conductance attributed to the Hofstadter spectrum were seen to arise when the BN lattice is nearly aligned with the graphene lattice~\cite{Dean2012, Ponomarenko2012}.

The nature of the conductivity at the charge-neutrality point $\sigmanp$ has been debated since the first graphene-based devices were fabricated. Theory for ballistic graphene predicts a value of $4e^2/\pi h$ for $\sigmanp$~\cite{Nilsson2006, Tworzydlo2006, Gorbar2002}. However, early experiments on graphene measured $\sigmanp$ from 2 to 12\,$\Gq$~\cite{Chen2008,Tan2007}. It was soon realized that $\sigmanp$ was sample-dependent, determined by the density of carriers in electron and hole puddles produced by static charges on or near the sheet of graphene~\cite{Adam2007}. In suspended graphene devices~\cite{Bolotin2008} the conductivity showed a more pronounced temperature dependence, but still saturated at low temperature and remained higher than $4e^2/\pi h$. Recently the potential landscape in graphene has been made artificially clean by screening potential fluctuations with a second nearby, doped graphene sheet~\cite{Ponomarenko2011, Kechedzhi2012}. In that work, instead of saturating at values near $\Gq$, $\sigmanp$ dropped with a power-law temperature dependence $T^\alpha$, where $\alpha$=2 for the most insulating samples, down to $T$=4\,K. Further, the authors observed a strong magnetoresistance in the temperature regime above 10\,K and attributed it to weak localization, inferring that ultra-clean graphene may be an Anderson insulator. Alternatively, it has been postulated~\cite{DasSarma2012} that this temperature dependence reflects increasing order causing the sample to become more insulating at low density.  Here $\sigmanp \propto T^{\alpha}$ naturally emerges as the temperature dependence of conduction through a landscape of electron and hole puddles. A complete understanding of this insulating behavior, so far appearing only in graphene on BN, is lacking.

\begin{figure}[t!]
\center \label{fig1}
\includegraphics[width=3 in]{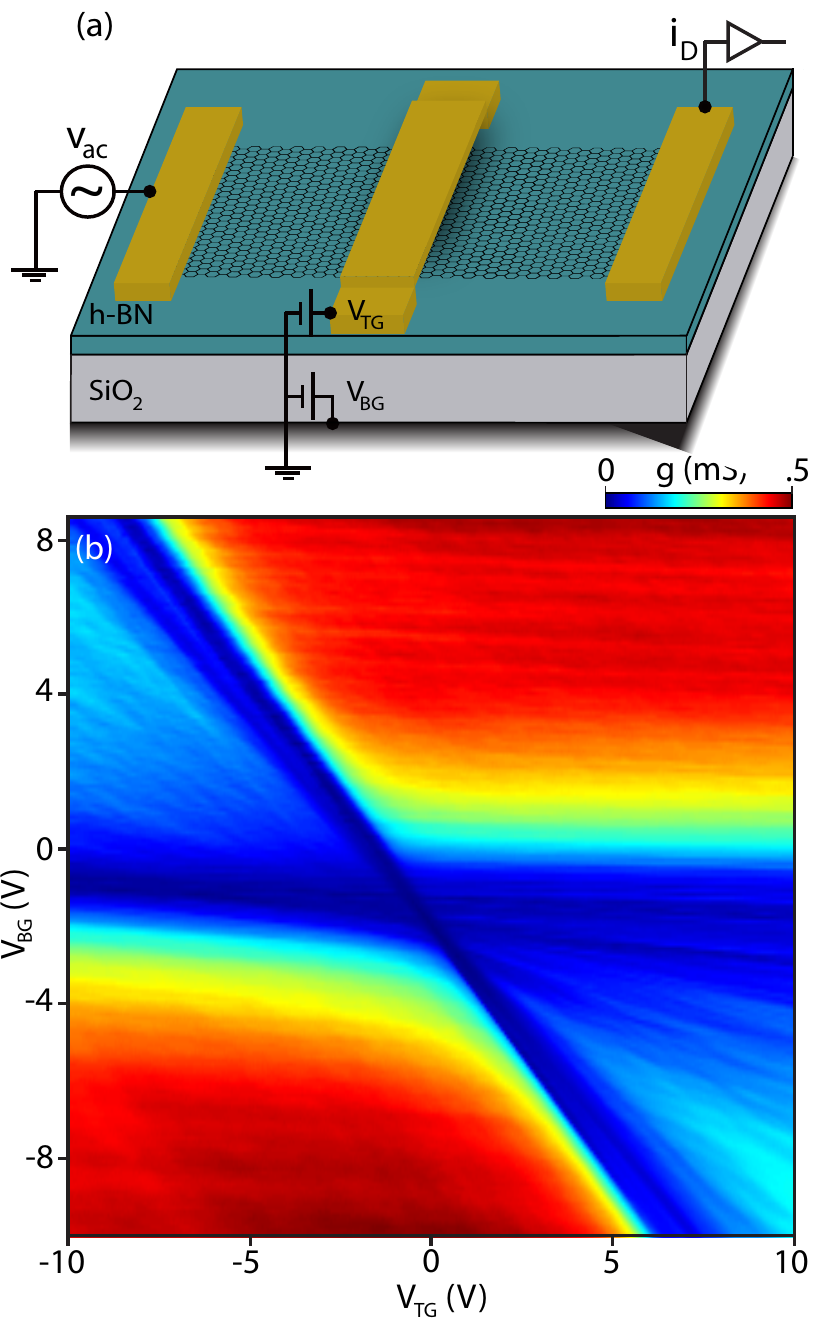}
\caption{\footnotesize{(a) Schematic of the device in voltage biased mode. A bias $v_{AC}$ is applied to the sample, current $i_{D}$ is collected at the drain and measured with a lock-in amplifier. A voltage $V_{BG}$ is applied to the degenerately-doped substrate to control the carrier density in the whole sheet. $V_{TG}$ is applied to a suspended metallic gate, 70\,nm above the graphene sheet, which varies the carrier density underneath it. (b) Conductance $g$ as a function of both gate voltages, measured at a temperature $T$=4\,K.}}
\end{figure}

In this Letter, we report on electronic transport in thirteen single-layer graphene devices, some with a top gate and some without. For the two devices where the top-gate is the closest to the graphene ($\sim$70\,nm), the conductivity at the charge neutrality point (CNP) decreases by more than 2 orders of magnitude with decreasing temperature, whereas the non top-gated samples show a more conventional temperature dependence. Here we focus on one top-gated device where the dependence of the conductance at the CNP ($\gnp$) on temperature ($T$) and perpendicular magnetic field ($B$) is investigated at temperatures down to 20\,mK. The temperature dependence is very strong down to $T$=400\,mK and can be fit to a power-law $\gnp \propto T^{\alpha}$ with $\alpha$\,$\approx$0.48$\pm 0.05$. Application of $B$ drives the system more insulating, where at $B$$\sim$100\,mT the $\nu$=0 state is entered, with no intermediate transition to the 2\,$\Gq$ quantum Hall plateau, indicating that a spin or valley (or both) symmetry is broken at very low fields. This direct transition to the $\nu$=0 state has not been observed before in graphene. We speculate on the origin of this effect.

We fabricated our devices using hexagonal-boron nitride (h-BN) as a substrate for graphene, with good electronic properties~\cite{Dean2009,Xue2011} resulting from the flatness and cleanliness of h-BN flakes. Details of the fabrication are described in Ref.~\cite{Suppinfo} and a schematic of the top-gated device geometry is shown in Fig.~1(a). Graphene devices were measured in two different cryostats: a variable-temperature insert enabling temperature-dependent transport measurements from 300\,K down to 1.7\,K, and a dilution fridge where samples were measured at lower temperatures, down to 20\,mK. The conductance $g$ is determined in a standard voltage-biased lock-in measurement with an excitation voltage of 4 $\mu$V at 92.3Hz. The resistance $r$ is defined as 1/$g$. DC voltages are applied to the top-gate ($\Vtg$) and back-gate ($\Vbg$). A table of all thirteen devices measured in this work, including mobility and resistance at the CNP ($\rnp=1/\gnp$) can be found in Ref.~\cite{Suppinfo}. 

$g$ is shown in Figure 1(b)  as a function of $\Vtg$ and $\Vbg$ at $T$=4\,K. The carrier density can be controlled independently and with either polarity underneath and outside the top-gated region. As in previous work on dual-gated graphene \cite{Williams2007, Huard2007}, $g$ exhibits local minima along two intersecting lines corresponding to each region being tuned through the CNP.  However, unlike in typical dual-gated graphene devices, $\gnp$ is much smaller than $\Gq$ along these lines. $g$ was also measured in a 4-probe geometry with similar results, ruling out poor contact resistance at the CNP as the source of this result.  Underneath the top-gate, $\Ctg\Vtg+\Cbg\Vbg=0$ (where $\Ctg$ is the top gate capacitance and $\Cbg$ is the back gate capacitance) at the CNP, yielding a top-gate-to-back-gate capacitance ratio of 1.3 from the slope of the diagonal line in the ($\Vbg, \Vtg$) plane. $\Cbg$ is 5.94($\pm$0.5)$\times$10$^{10}$cm$^{-2}$V$^{-1}$, as extracted from the periodicity of Shubnikov-de-Haas oscillations. Using a parallel-plate capacitance model, we estimate that the top gate is 70\,nm away from the flake, which was confirmed by atomic force microscopy. The device exhibits little intrinsic doping, with a CNP voltage of -1\,V on the back-gate, and a mobility of 70,000\,cm$^{2}$/Vs, as extracted from a linear fit to $g$($\Vbg$) at the CNP. We note that the mobility of top-gated and non top-gated regions was comparable, which shows that the suspended gate does not degrade the electronic properties of our device. 

\begin{figure}[t!]
\center \label{fig1}
\includegraphics[width=3 in]{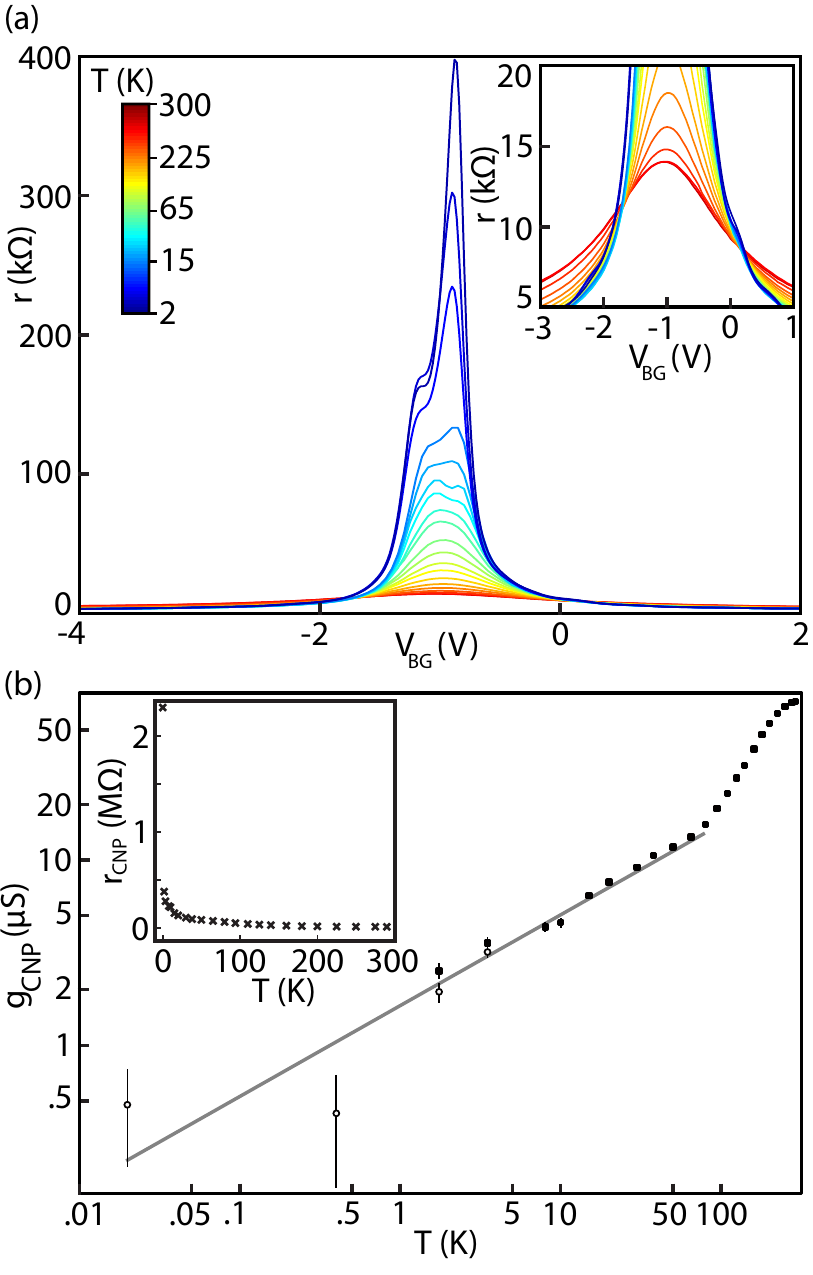}
\caption{\footnotesize{(a) $r$($V_{BG}$) at temperatures ranging from 300\,K to 2\,K. Inset: close-up view of the intersection of the curves (b) $\gnp$($T$) with open circles and filled squares were measured in the dilution fridge and variable temperature cryostat, respectively. Error bars correspond to one standard deviation: for T$>$5\,K, these are smaller than the dots. The grey line corresponds to a fit $\gnp\propto T^{\alpha}$, with $\alpha=0.48\pm0.05$ for T$<$80\,K. At higher temperature $\gnp$ rises with a faster exponent. Inset: resistance at the neutrality-point $\rnp$ as a function of temperature.}}
\end{figure}

Unlike typical graphene samples, $r$ as a function of $\Vbg$ has a strong temperature dependence in our device, as shown on figure 2 for $\Vtg$=0. $\rnp$ dramatically increases at low temperature, from 13\,k$\Omega$ at $T$=300\,K to 400\,k$\Omega$ at $T$=2\,K [Fig. 2(a)]. By contrast, $\rnp$ for all devices without a top gate is typically around 10\,k$\Omega$ at 2\,K~\cite{Suppinfo}, including for a separate device made in the same sheet of graphene as the device shown on Fig. 2 but without a top-gate. This is comparable to what is commonly seen in good-quality, single-layer graphene devices. The top-gated sample has an insulating temperature dependence close to the CNP, for -1.8\,V$\le \Vbg \le$0\,V, whereas for higher carrier densities it is metallic [inset, Fig. 2(a)].

$\rnp$ was measured at lower temperature in a separate cool-down using a dilution fridge: further lowering $T$ to 400\,mK increases $\rnp$, at which point it measures 2.3\,M$\Omega$ [inset, Fig. 2(b)], then remains constant down to 20\,mK within experimental error. Fig. 2(b) shows $\gnp$($T$), which follows a power law $\gnp\propto T^{\alpha}$ as a function of the temperature, with $\alpha$=0.48$\pm$0.05. This insulating behavior is not due to the opening of a hard band gap, which would lead to an exponentially-activated conductivity. The temperature dependence is also slower than the $T^{2}$ dependence measured in Ref.~\cite{Ponomarenko2011} (and expected from the Boltzmann equation with electron-electron scattering). A temperature dependence similar to ours was reported in suspended graphene~\cite{Bolotin2008}, although the overall conductance was several orders of magnitude higher and the device was not top-gated. 

\begin{figure}[t!]
\center \label{fig1}
\includegraphics[width=3 in]{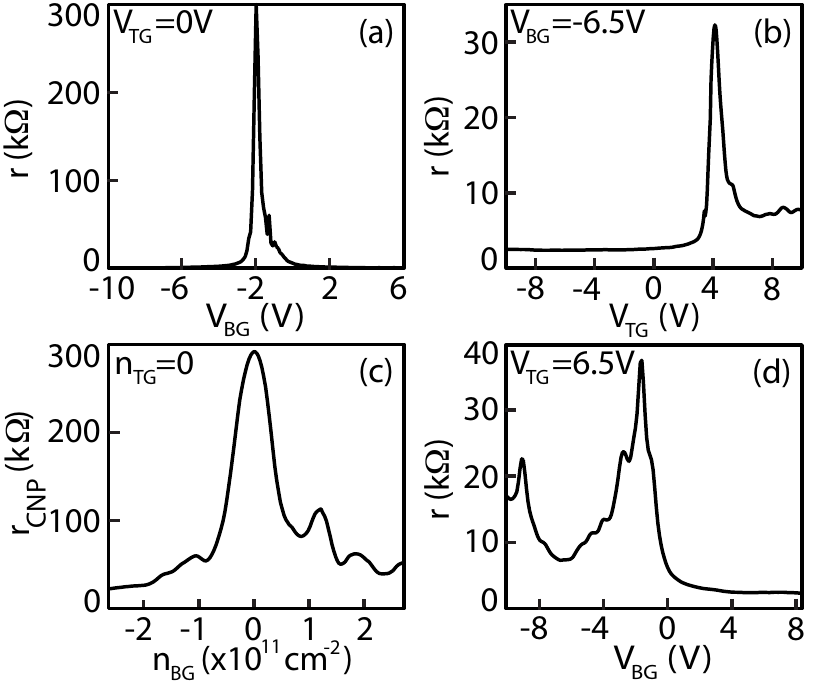}
\caption{\footnotesize{Cuts of the resistance $r$ measured at T=4K. (a) r($\Vbg$, $\Vtg$=0). (b) r($\Vtg$, $\Vbg$=-6.5V). (c) r(n$_{BG}$, n$_{TG}$=0). (d) r($\Vbg$, $\Vtg$=6.5V)}}
\end{figure}

The peak resistance depends on densities both under and outside the top-gate, as shown in Fig. 3 at $T$=4\,K. The resistance when the density is uniform - for $\Vtg$=0\,V - is shown in Fig. 3(a). When the density outside the top-gated region is nonzero [Fig. 3(b)], $r$($\Vtg$) shows the electron-hole asymmetry reported in Ref.~\cite{Huard2007,Williams2007}. The resistance when the top-gated part of the device is at the neutrality-point ($n_{TG}$=0) is shown in Fig. 3(c), as a function of the carrier density outside the top-gated region ($n_{BG}$). The device has the largest resistance at double-neutrality ($n_{BG},n_{TG} \sim$ 0), in contrast with bilayer graphene, where a similar plot would produce the largest value of $\rnp$ at large n$_{BG}$, associated with gap opening by a large transverse electric field~\cite{Castro2007}. The peak resistivity steeply decreases from 300\,k$\Omega$ at double neutrality to less than 25\,k$\Omega$ when the carrier density outside the top-gated region is more than 2.5 $\times$10$^{11}$cm$^{-2}$. The local maximum of r($\Vtg$,$\Vbg$) corresponding to charge-neutrality outside the top-gated region yields $\rnp \sim$32k$\Omega$ per square for non-top-gated graphene, which means that the top-gated part of the device accounts for most of the 300 k$\Omega$ measured at double neutrality.

A low-field fan diagram [$g$($\Vbg, B$)] is measured at $T$=2\,K and shown in Fig.~4(a). Away from the CNP, we observe plateaus for $\nu=2,6,10$ (dashed lines) that are well-developed on the hole side for $B$$>$0.5T. The cut g($\Vbg, B$=1\,T) shows these plateaus in addition to the $\nu$=0 plateau around the CNP [Fig. 4(c)]. Interestingly, the  broken-symmetry $\nu$=0 state seems to persist all the way down to very low fields [dark blue region that runs between $\Vbg\sim$0 and -1V in Fig.~4(a)]. The minimum conductance $\gnp$ as a function of $B$ is shown in Fig.~4(b). The value of $\Vbg$ at which the minimum occurs shifts slightly upward with magnetic field, drifting by less than 0.1V as $B$ is increased. $\gnp$($B$) monotonically decreases on a scale of $\sim$100\,mT and enters the $\nu$=0 gap without first transitioning to the 2$\Gq$ plateau.

\begin{figure}[t!]
\center \label{fig1}
\includegraphics[width=3 in]{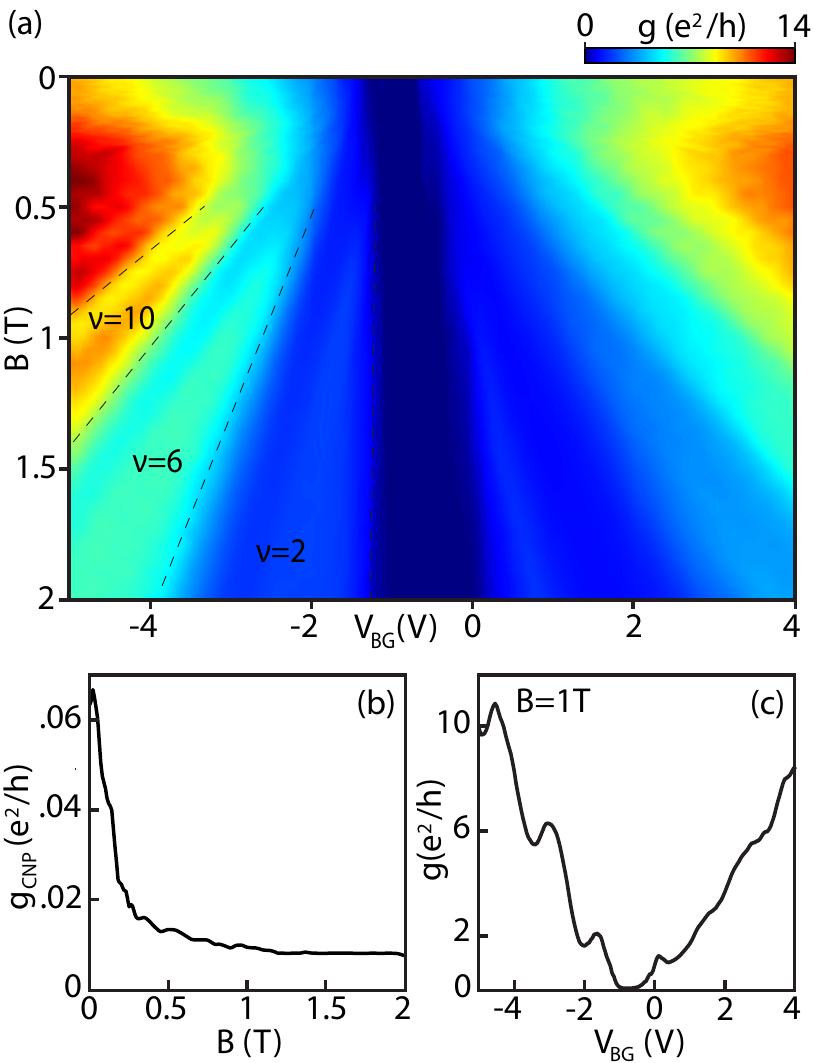}
\caption{\footnotesize{(a) $g$($B,\Vbg$) measured at T=2\,K and in a two-probe configuration. The transitions between plateaus are underlined for $\nu=$0, 2, 6, 10. (b) Minimum conductance as a function of the magnetic field $\gnp$(B), showing a direct transition to the $\nu$=0 state as a function of $B$. (c) Cut of g($V_{BG}$) at B=1\,T.}}
\end{figure}

\emph{Discussion}. The temperature dependence of the peak resistance in graphene is often used to estimate the disorder level in a sample~\cite{Bolotin2008}. Cleaner samples show a slightly higher peak resistivity which increases as $T$ is lowered, and saturates when $k_{B}T$ is smaller than the Fermi energy's fluctuations $\dEf$. However, this interpretation seems insufficient to explain our data: the temperature dependence in our device is very strong down to $T$$\approx$ 400\,mK  which would translate to surprisingly small Fermi energy fluctuations $\dEf \approx 40$\,$\mueV$ -- corresponding to density fluctuations of less than $10^{6}$ cm$^{-2}$ -- two orders of magnitude lower than in most reported suspended devices~\cite{Mayorov2012}. 

The presence of a top-gate is likely to contribute to this insulating behavior, since we have yet to observe this effect in non-top-gated devices, and screening by the top-gate could explain reduced density fluctuations. As charge puddles get shallower and further apart, the minimum conductivity is expected to drop as it becomes harder for electrons to percolate~\cite{Cheianov2007,Adam2007}. However, the top gate is tens of nanometers away from the flake, significantly further than reported in Ref.~\cite{Ponomarenko2011}, and it is therefore surprising that the screening should have such dramatic consequences, as interactions should naively be screened only on distances larger than the spacing to the top gate. 

It was recently proposed that in graphene on h-BN devices, electron interactions could enhance the substrate-induced valley-dependent pseudo-potential, hence breaking the valley-symmetry~\cite{Giovannetti2007, Levitov2013, Kindermann2012}. This gap has not been observed yet, but could explain a diverging resistance at the Dirac point in very clean samples. Our low-field magneto-transport data suggest that valley symmetry-breaking may indeed play a role in this insulating behavior. The conductance around the charge neutrality point is never quantized and always very small compared to 2\,$\Gq$ [Fig. 4(b)]. The monotonic decrease of $\gnp$ indicates that the insulating behavior observed at zero field becomes stronger as the magnetic field breaks the valley symmetry and the $\nu$=0 gap opens up. This is qualitatively different from what we and others~\cite{Suppinfo,Young2012} observed at the CNP for non-top-gated devices: the two-terminal conductance usually develops plateaus at a moderate magnetic field ($\sim$1-2\,T). In this regime, $\gnp$ depends on the aspect-ratio of the device~\cite{Williams2009} but is always on order $\Gq$ (corresponding to $\nu$=$\pm$2). We usually observe that $\gnp$ decreases from $\sim\Gq$ to zero on a scale of several Tesla, as the valley degeneracy of the $n$=0 Landau level is lifted~\cite{Zhang2006, Suppinfo}.

The very steep decrease in the conductance on Fig. 4(b), B=100mT at half-maximum, suggests that either the Landau level broadening is extremely small, or the valley-dependent interactions grow more rapidly with B than in regular devices. In a magnetic field, the energy of Coulomb interactions is on the order $E_{C}= \frac{e^{2}}{\epsilon l_{B}}$, where $l_{B}$ is the magnetic length $l_{B}$=$\sqrt{\frac{\hbar}{eB}}$. While to first order these interactions preserve valley symmetry, it has been shown that higher-order terms break this symmetry and are on order $\delta E_{C}=\frac{a}{l_{B}}E_{C}$, where $a$ is the spacing between neighboring carbon atoms~\cite{Young2012}. A naive estimate of this contribution gives $\delta E_{C}$$\sim$1\,meV/T. Interestingly, the field dependence of the $\nu$=0 gap has been studied in Ref~\cite{Young2012}, where it was found that the effective $g$ factor $g_{\Delta_ {0}}\equiv d\Delta_{o}/dB$ for the $\nu=0$ state had this same order of magnitude. Using this result gives a naive upper-bound of $100 \mueV$ for the Landau level broadening due to the Fermi level fluctuations, in good agreement with our previous estimate based on the low-temperature saturation of $\gnp$($T$). 

However, the cuts of the resistance shown on Fig.~3 suggest that $r$ under the top-gate doesn't depend solely on n$_{TG}$. Otherwise, $\rnp$ would only decrease by $\sim$60 k$\Omega$, the peak resistance of the non-top-gated part of the device, when the two non-top-gated squares of the device are tuned away from charge neutrality, instead of the wide range observed in Fig. 3(c). Due to the roughness of the top-gate, its geometric capacitance is not perfectly uniform, which induces stronger density fluctuations at higher transverse electric field than at double-neutrality. The full width at half-maximum of  $\rnp$($\Vtg$) is often used to estimate the disorder level in a sample. At 4K, the width is only $\delta n\sim$2.3x10$^{10}$cm$^{-2}$ at double neutrality but $\delta n\sim$1.3x10$^{11}$cm$^{-2}$ for $\Vbg$=-10V. If the insulating behavior we observe is indeed due to the extreme cleanliness of the top-gated part of the device, it is understandable that $\rnp$ decreases at higher gate voltages due to stronger density fluctuations. It is also possible that the interfaces between the top- and non-top-gated regions play a role in the diverging resistance we measured. The resistance of an interface between regions of different densities is not expected to be so high~\cite{Cheianov2008}, but the interface resistance between two states with different charge correlations might be substantial, and this could come into play if a new state could exist due to screening underneath the top-gate.

We thank A. J. Bestwick, C. R. Dean, P. Gallagher, P. Jarillo-Herrero, M. Y. Lee, L. S. Levitov, J. Sanchez-Yamagishi for useful technical help and fruitful discussions. This work was funded by the Center on Functional Engineered Nano Architectonics (FENA), the W. M. Keck Foundation and the Stanford Center for Probing the Nanoscale, an NSF NSEC, supported under grant No. PHY-0830228. F. Amet acknowledges support from a Stanford Graduate Fellowship.

\clearpage

\section{Supplementary information}
\subsection{Device fabrication}

Polyvinyl alcohol (2$\%$ in water) is spun at 6000\,rpm on bare silicon and baked at 160\,$^{\circ}$C for 5\,mins, resulting in a 40\,nm thick layer. Then, a layer of PMMA (5$\%$ in anisole) is spun at 2400\,rpm and baked 5\,mins at 160\,$^{\circ}$C. The total polymer thickness is on the order of 450\,nm. Graphene is then exfoliated on this stack using Nitto tape, located using optical microscopy, and characterized with Raman spectroscopy. Boron nitride flakes are exfoliated on a silicon wafer piece with a 300\,nm thick thermal oxide, then baked in flowing Ar/O$_{2}$ at 500\,$^{\circ}$C for 8 hours to remove organic contamination such as tape residue~[S1]. The cleanliness and thickness of the flakes are characterized with atomic force microscopy and Raman spectroscopy. 

The PVA layer is dissolved in deionized water at 90\,$^{\circ}$C, which lifts-off the PMMA membrane with the graphene attached to it. The membrane is then adhered across a hole in a glass slide~[S2] and baked at 110\,$^{\circ}$C. We then use a home-made probe-station to align the graphene flake on top of the boron nitride substrate. Once both flakes are in contact, the stack is baked at 120\,$^{\circ}$C to promote adhesion. The PMMA layer is dissolved in hot acetone, then rinsed in IPA, which leaves the graphene flake on top of the boron nitride flake. This stack is annealed in flowing Ar/O$_{2}$ at 500\,$^{\circ}$C for 4 hours, which removes process residue and leaves the graphene flake pristine, as checked by Raman spectroscopy. 

We use regular electron beam lithography combined with oxygen plasma etching to pattern a graphene Hall bar. In order to fabricate a suspended top gate above the device~[S3-6], the samples are spin-coated at 6000\,rpm with a solution of polymethyl-methacrylate (950k), 3$\%$ in anisole, then baked at 160\,$^{\circ}$C for 5 minutes. An additional layer of methyl-methacrylate  (8.5$\%$ in ethyl lactate) is spin-coated at 6000\,rpm and baked 5 minutes at 160\,$^{\circ}$C. The MMA layer is 50$\%$ more sensitive to electron irradiation than the PMMA layer and it is therefore possible to develop the top resist layer without exposing the bottom layer.  The e-beam writing system we used is a JEOL 6300, with an acceleration voltage of 100\,keV. The contacts and the feet of the suspended bridge are exposed with 1000\,$\mu \mathrm{C/cm}^{2}$, which is enough to dissolve both resist layers upon development in MIBK/IPA (1:3) for 45\,sec. The span of the suspended bridge is only exposed with a base dose of 650\,$\mu \mathrm{C/cm}^{2}$, which only develops the top resist layer. After development, the device is cleaned for 2 minutes with UV ozone, then metallized with 1nm of chromium and 200\,nm of gold. 

Fig. 5 is a scanning electron micrograph of a graphene on boron-nitride device with a suspended top-gate that has been fabricated using the same recipe. Suspended gates as long as 7 microns have been fabricated using this recipe. These usually resist further heat treatment as well as cryogenic temperatures. We found that gate voltages as high as 40\,V can be applied to the top-gate without damaging it. 

\begin{figure}[t!]
\center
\includegraphics[width=3 in]{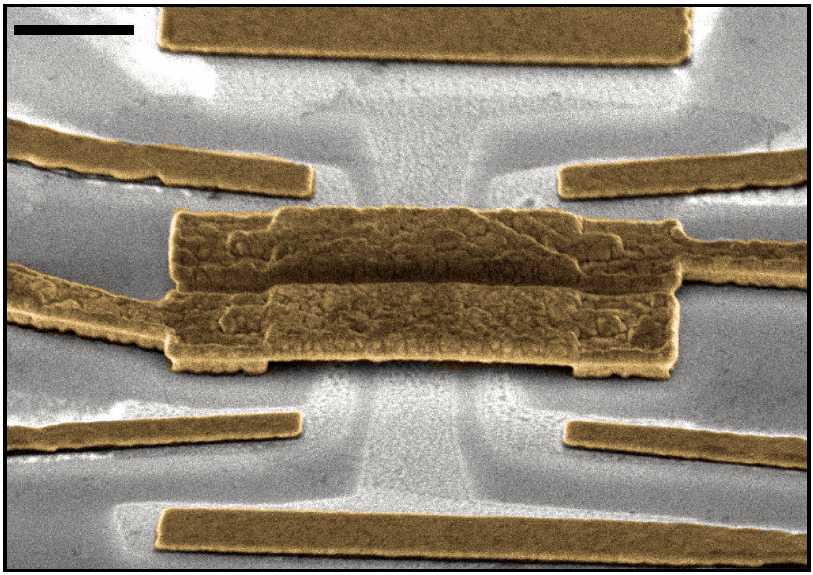}
\caption{\footnotesize{Scanning electron micrograph of a graphene on boron nitride device with a suspended top gate. Scale bar=1 $\mu$m. }}
\label{figS1}
\end{figure}

\begin{figure}[t!]
\center 
\includegraphics[width=3 in]{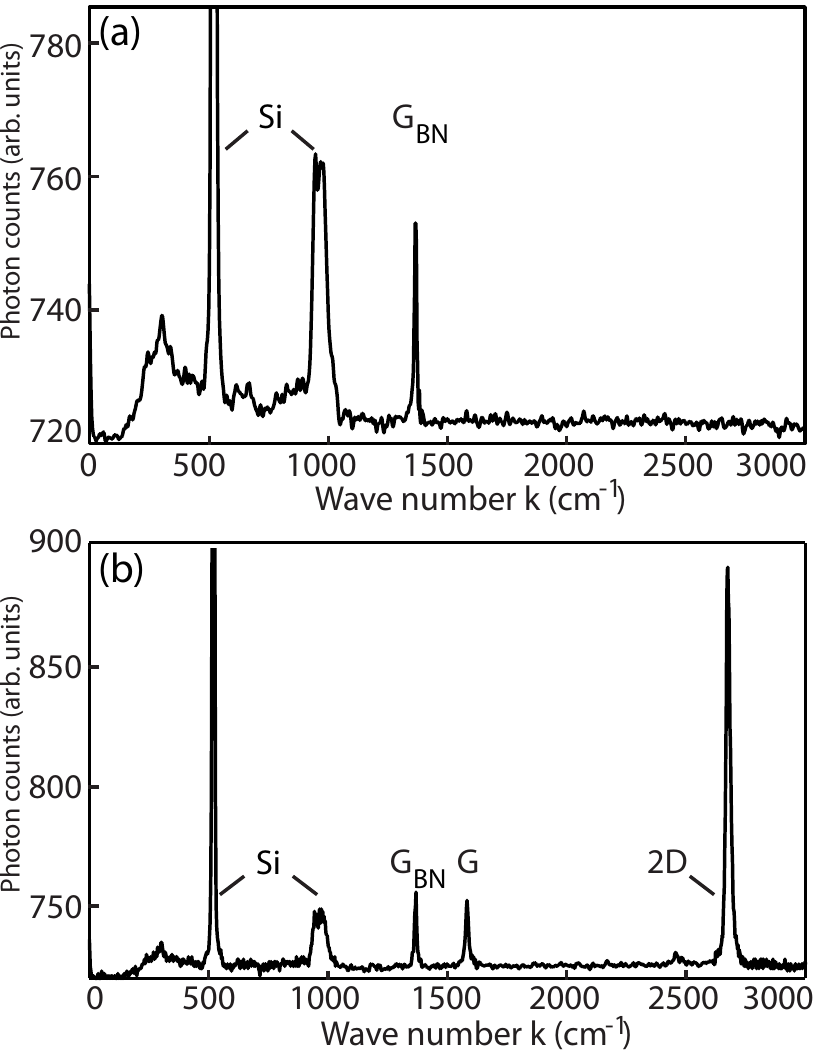}
\caption{\footnotesize{(a) Raman spectrum of the boron nitride flake used for the device described in the main paper, prior to transferring the graphene flake. The boron nitride Raman peak is labelled  $G_{BN}$.  (b) Raman spectrum of the whole graphene on boron nitride device after oxygen annealing and before contacts were made to the flake. The peaks labelled G and 2D are attributed to graphene.}}
\label{figS2}
\end{figure}

\subsection{Raman spectra}

We check the cleanliness of every boron nitride flake we use with atomic force microscopy and Raman spectroscopy.  Fig. 6(a) shows the Raman spectrum of the boron nitride flake used in the device studied in the paper, prior to transferring the graphene flake. The absence of a broad background signal~[S2] is a very clear indication that the flake is free of any kind of organic contamination.  The Raman spectrum of the transferred device before the last lithographic step -when the suspended gate is patterned, is shown on Fig. 6(b). The ratio of the amplitudes of the 2D and G peak is $\mathrm{I}_{\mathrm{2D}}/\mathrm{I}_{\mathrm{G}}=5.5$, and the full half width of the 2D peak is 19 cm$^{-1}$, indicating with no ambiguity that the flake studied here is single-layer graphene~[S3]. The absence of a broad background signal attests for the cleanliness of the device.

\subsection{Resistance of a non top-gated sample}

A typical resistivity of a non top-gated part of the sample described in the main paper is shown on Fig. 7, measured at T=4\,K. The peak resistivity is $\sim$10k$\Omega$ and does not diverge as the temperature is lowered.

\begin{figure}[t!]
\center 
\includegraphics[width=3 in]{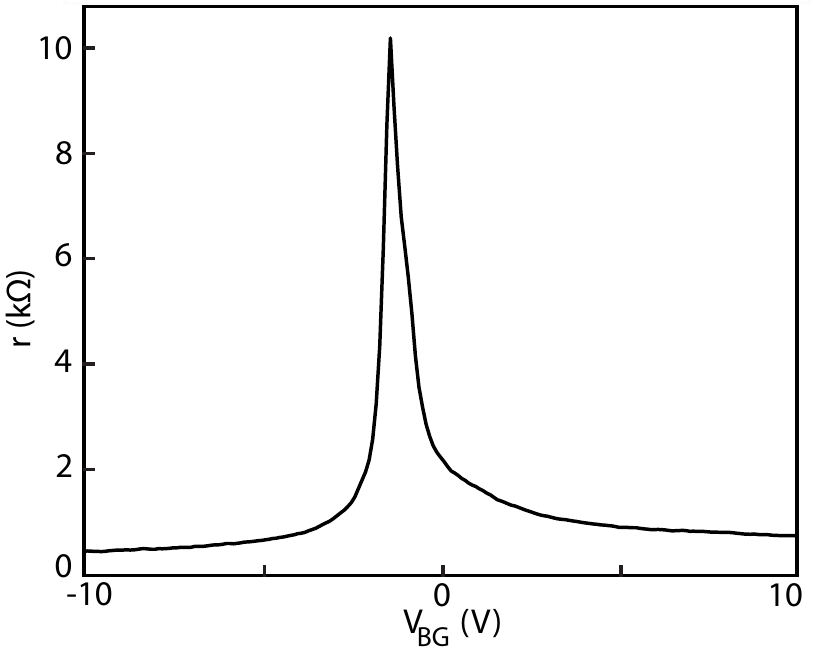}
\caption{\footnotesize{Resistivity as a function of the $\Vbg$ for the control device}}
\label{figS3}
\end{figure}

\subsection{Quantum Hall conductance of a non top-gated device}

\begin{figure}[t!]
\center
\includegraphics[width=3 in]{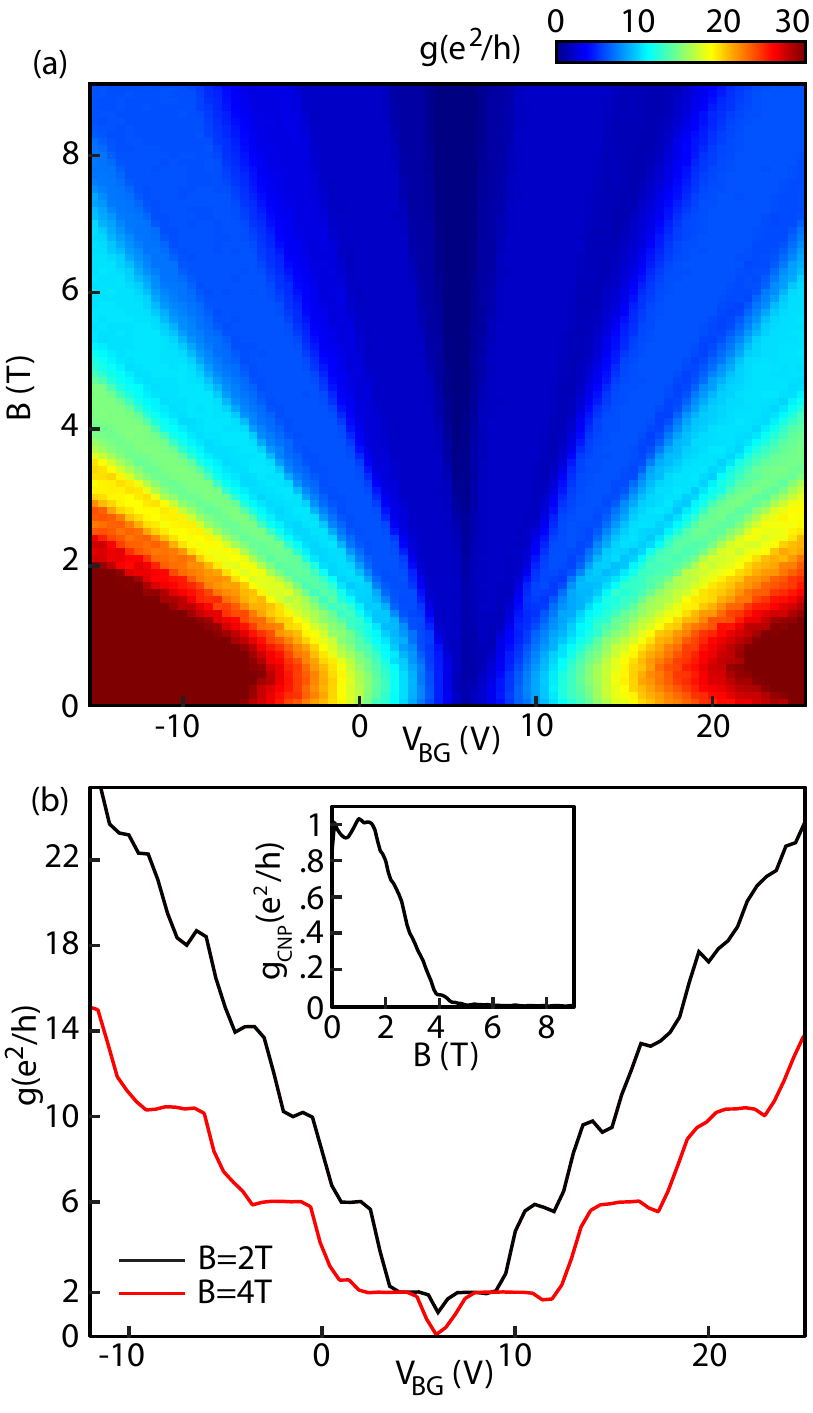}
\caption{\footnotesize{(a) Two terminal inductance as a function of the perpendicular magnetic field and the back-gate voltage. (b) Cuts of the conductance as a function of the back gate voltage at $B$=2\,T (black line) and $B$=4\,T (red line). Inset: $\gnp$(B) for this sample.}}
 \label{figS4}
\end{figure}

The two-terminal quantum Hall conductance of a graphene on boron nitride device with no top-gate is shown on Fig. 8, the mobility for this sample is 45 000 cm$^{2}$/Vs. The conductance is measured at T=2\,K in a two-probe configuration. Most of the measurements in this paper use this geometry, as voltage-biased measurements of the conductance are more convenient than current-biased ones when the device is insulating, around the charge neutrality point. Quantum Hall plateaus are better resolved on the hole side than on the electron side, which we found to be common for two-terminal quantum Hall conductance measurements on other graphene on boron nitride devices. 

As the magnetic field increases, the degeneracy of each Landau level splits and we observe new quantized plateaux. In particular, the $\nu$=0 phase slowly appears around $B$=3\,T, as seen on the inset Fig. 8. This is in sharp contrast with the abrupt transition seen in figure 4b of the main paper, which occurs at a much lower field. Fig. 8(b) shows two sections of the fan diagram at $B$=2\,T and $B$=4\,T. The conductance at 2\,T shows the standard sequence of plateaux, as seen in other two terminal devices with no splitting of the Landau levels~[S4]. We stress that the small dip in conductance at the neutrality point at 2\,T is only an artifact from the two terminal geometry, as shown in Ref.~[S4], and is not related to the opening of the $\nu$=0 gap. At 4\,T, the device starts being insulating around the charge neutrality point, as the $\nu$=0 gap opens up, and the other plateaus become further resolved.

\section{Table of peak resistances}

We show in Tables I and II the peak resistances $\rnp$ and mobilities of the devices included in this study. $\rnp$ was always significantly smaller than $h/e$$^{2}$ for non top-gated devices. The insulating behavior at the CNP is most pronounced for the two devices with the highest top-gate capacitance, with $\rnp >>$ $h/e$$^{2}$. The two devices marked with a star were patterned in the same graphene flake.

\begin{table}[ht]
\caption{Measured properties of top-gated devices.}
\centering
\begin{tabular}{c c c}
\hline\hline Carrier Mobility \\ (cm$^2$/Vs) & $\rnp$ ($k\Omega$/sq) &
$\Ctg$ (nF/cm$^2$) \\ [0.5ex]
\hline
70,000* & 760 & 12.8 \\
6,000 & 45 & 12.8 \\
120,000 & 14 & 9.8 \\
45,000 & 8.5 & 3.3 \\
35,000 & 8.9 & 3.3  \\
15,000 & 7.5 & 2.8  \\
[1ex] \hline\hline
\end{tabular}
\label{table:nonlin}
\end{table}

\begin{table}[ht]
\caption{Measured properties of non-top-gated devices.}
\centering
\begin{tabular}{c c}
\hline\hline Carrier Mobility \\ (cm$^2$/Vs) & $\rnp$ ($k\Omega$/sq) \\ [0.5ex]
\hline
85,000 & 16 \\
70,000* & 11 \\
70,000 & 12 \\
62,000 & 8.8 \\
45,000 & 8.9 \\
45,000 & 8.8  \\
40,000 & 6.8  \\
40,000 & 5.2  \\
[1ex] \hline\hline
\end{tabular}
\label{table:nonlin}
\end{table}

\section{Supporting references}
\begin{itemize}
\item[S1. ]
A. G. F. Garcia, M. Neumann, F. Amet, J. R. Williams, K. Watanabe, T. Taniguchi and D. Goldhaber-Gordon, Nano Lett. 12(9), 4449-4454 (2012).

\item[S2. ]
C. Dean, A. F. Young, I. Meric, C. Lee, L. Wang, S. Sorgenfrei, K. Watanabe, T. Taniguchi, P. Kim, K. L. Shepard and J. Hone, Nature Nano. \textbf{5}, 722Ð726 (2010).

\item[S3. ]
J. Velasco Jr, G. Liu, W. Bao and C. N. Lau, New J. Phys. \textbf{11}, 095008 (2009).

\item[S4. ]
R. T. Weitz, M. T. Allen, B. E. Feldman, J. Martin and A. Yacoby, Science \textbf{330}, 812 (2010).

\item[S5. ]
M. T. Allen, J. Martin and A. Yacoby, Nature Comm. \textbf{3}, 934 (2012).

\item[S6. ]
R. V. Gorbachev, A. S. Mayorov, A. K. Savchenko, D. W. Horsell and F. Guinea, Nano Lett. 8(7), 1995-1999 (2008).

\item[S7. ]
A. C. Ferrari, J. C. Meyer, V. Scardaci, C. Casiraghi, M. Lazzeri, F. Mauri, S. Piscanec, D. Jiang, K. S. Novoselov, S. Roth and A. K. Geim, Phys. Rev. Lett. \textbf{97}, 187401 (2006). 

\item[S8. ]
J. R. Williams, D. A. Abanin, L. DiCarlo, L. S. Levitov and C. M. Marcus, Phys. Rev. B \textbf{80} 045408 (2009).
\end{itemize}

\end{document}